\documentclass[twocolumn]{aastex6}
\usepackage{amsmath}

\slugcomment{Accepted to ApJ}

\shorttitle{Color and Stellar Mass Dependent Clustering}
\shortauthors{Law-Smith \& Eisenstein}

\begin{document}

\title{The Color and Stellar Mass Dependence \\of Small-Scale Galaxy Clustering in SDSS-III BOSS}

\author{Jamie Law-Smith\altaffilmark{1} and Daniel J. Eisenstein\altaffilmark{2}}
\altaffiltext{1}{Department of Astronomy and Astrophysics, University of California Santa Cruz, CA 95064, USA; \href{mailto:lawsmith@ucsc.edu}{lawsmith@ucsc.edu}}
\altaffiltext{2}{Center for Astrophysics, Harvard University, 60 Garden St., Cambridge, MA 02138, USA}

\begin{abstract}
We measure the color and stellar mass dependence of clustering in spectroscopic galaxies at $0.6 < z < 0.65$ using data from the Baryon Oscillation Spectroscopic Survey component of the Sloan Digital Sky Survey. We greatly increase the statistical precision of our clustering measurements by using the cross-correlation of 66,657 spectroscopic galaxies to a sample of 6.6 million fainter photometric galaxies. The clustering amplitude $w(R)$ is measured as the ratio of the mean excess number of photometric galaxies found within a specified radius annulus around a spectroscopic galaxy to that from a random photometric galaxy distribution. We recover many of the familiar trends at high signal-to-noise ratio. We find the ratio of the clustering amplitudes of red and blue massive galaxies to be $w_\text{red}/w_\text{blue} = 1.92 \pm 0.11$ in our smallest annulus of 75--125 kpc. At our largest radii (2--4 Mpc), we find $w_\text{red}/w_\text{blue} = 1.24 \pm 0.05$. Red galaxies therefore have denser environments than their blue counterparts at $z \sim0.625$, and this effect increases with decreasing radius. Irrespective of color, we find that $w(R)$ does not obey a simple power-law relation with radius, showing a dip around 1 Mpc. Holding stellar mass fixed, we find a clear differentiation between clustering in red and blue galaxies, showing that clustering is not solely determined by stellar mass. Holding color fixed, we find that clustering increases with stellar mass, especially for red galaxies at small scales (more than a factor of 2 effect over 0.75 dex in stellar mass).
\end{abstract}


\section{Introduction}
A core goal of extragalactic astronomy is to understand the impact of the history of galactic dark matter halos and environments on galaxy properties. One way to do this is with detailed studies of galaxy clustering as a function of galactic observables (such as position, redshift, color, luminosity, and stellar mass). If richer galaxy clustering occurs in regions of denser dark matter, we can learn about the relationship between galaxy properties and large-scale structure from the relationship between galaxy properties and clustering.
	
Besides helping us better understand the relationship between galaxies and halos, this work can shed light on galaxy formation and evolution. Clustering dependence on galaxy properties has been studied at low redshifts ($z < 0.25$) with the original Sloan Digital Sky Survey \citep[SDSS;][]{Zehavi:2002, Zehavi:2005, Zehavi:2011}. It has been investigated at intermediate redshifts of $0.2 < z < 1$ with COSMOS \citep{Meneux:2009, Leauthaud:2012}, $0.2 < z < 1.2$ with CFHTLS \citep{McCracken:2008, Coupon:2012, Coupon:2015}, $0.5 < z < 1.1$ with the VIMOS Public Extragalactic Redshift Survey \citep[VIPERS;][]{Marulli:2013}, and $0.2 < z < 1.0$ with PRIMUS \citep{Skibba:2014, Skibba:2015}, though with smaller data sets and different methods than we use here. Galaxy clustering has been explored at $z\sim1$ with the DEEP2 Galaxy Redshift Survey \citep{Coil:2008, Mostek:2013} and the VIMOS-VLT Deep Survey \citep[VVDS;][]{Meneux:2006, Meneux:2008, Pollo:2006}, out to $z\sim2$ with UltraVISTA \citep{McCracken:2015}, and out to $z\sim3$ with the UKIDSS Ultra Deep Survey \citep{Hartley:2010, Hartley:2013}. In selecting intermediate redshifts of $0.6 < z < 0.65$, we can further quantify how the relationship between clustering and properties evolves over time. This work can also be compared with recent efforts using the same data sets and redshift range \citep{Guo:2013, Guo:2014}, as we use a different method (cross-correlation rather than auto-correlation).

Clustering dependence on galaxy properties is understood as a result of the density--morphology relation, where an increasing elliptical and S0 population and a corresponding decrease in spirals are associated with increasing cluster density \citep{Dressler:1980}. SDSS offers the largest statistical galaxy sample to date, enabling us to investigate more specific questions, such as the relationship between clustering and color or luminosity. With an early SDSS spectroscopic sample, clustering was found to depend on luminosity for red galaxies but not for blue galaxies \citep{Hogg:2003}. In addition, red galaxies were found to inhabit highly over-dense regions. For luminous red galaxies, a very strong luminosity--clustering dependence was found: up to a factor of 4 in clustering amplitude over a factor of 4 in luminosity \citep{Eisenstein:2005}. With the completion of the original SDSS survey, we obtained more definitive results on the clustering dependence on color and luminosity for $z < 0.25$ \citep{Zehavi:2002, Zehavi:2005, Zehavi:2011}. Clustering was found to increase slowly with luminosity for $L < L_*$ and more rapidly for higher luminosities; clustering was also found to have a continuously increasing trend with color from blue to red. These trends agree with earlier findings at similar redshifts using the 2dF Galaxy Redshift Survey \citep{Norberg:2001, Norberg:2002}. Studies at $0.2 < z < 1$ using zCOSMOS found a weak dependence of galaxy clustering on luminosity, and a mild dependence on stellar mass, particularly evident in the $0.5 < z < 0.8$ bin on small scales \citep{Meneux:2009}. Clustering of VIPERS galaxies at $0.5 < z < 1$ was found to monotonically increase with $B$-band magnitude and stellar mass \citep{Marulli:2013}. Studies using DEEP2 galaxies at $z\sim1$ \citep{Coil:2008} also found similar trends, though with no changes of the clustering within the red sequence. Studies using PRISM galaxies at $0.2 < z < 1$ found that red galaxies have stronger small-scale clustering compared to blue galaxies, as well as a strong color-dependent clustering within the red sequence \citep{Skibba:2014}.

With the SDSS-III BOSS (Baryon Oscillation Spectroscopic Survey) data, the same clustering studies were made in the redshift range $0.43 < z < 0.7$ \citep{Guo:2013}. Clustering was found to depend clearly on luminosity and color, with more luminous and redder galaxies generally exhibiting stronger clustering. This clustering dependence on luminosity and color has been interpreted using halo occupation distribution (HOD) models \citep[e.g.,][]{Zehavi:2005}. The HOD model fits the correlation results better than a power law and explains inflections at larger radii as a transition from one- to two-halo regimes. The HOD model explains the color dependences of clustering: less massive halos have blue central galaxies, while more massive halos have red central and satellite galaxies. The fraction of blue central galaxies decreases with increasing luminosity and host halo mass \citep{Zehavi:2005}.

In this paper we examine how the clustering of galaxies (i.e., the density of neighboring galaxies) scales with color and mass. We do this by cross-correlating spectroscopic and photometric data sets from the BOSS component of SDSS-III, as explained in the next section \citep[see also][]{Eisenstein:2003}. This technique has been used in several galaxy clustering studies to date \citep[e.g.,][]{Davis:1978, Yee:1987, Lilje:1988, Saunders:1992, Hogg:2003, Eisenstein:2005, Tal:2012, Tal:2013, Bray:2015}. We focus on cross-correlations as the spectroscopic data sets are limited in size and subdividing them into many subcategories leaves them starved for statistical precision. This is particularly true for the massive galaxies in BOSS. By introducing a color-selected set of fainter galaxies, we can use cross-correlations of small spectroscopic populations relative to the larger control sample. Although we lack the second spectroscopic redshift, the angular correlations are dominated by clustering of nearby pairs. On small enough angular scales, the increase in noise due to interlopers is smaller than the opportunity to use the denser tracer population.

This paper is organized as follows. In Section \ref{sec:methods} we describe the spectroscopic and imaging samples, their division to specific subsamples, and our cross-correlation method for measuring the correlation function. In Section \ref{sec:results} we present the clustering measurements and detailed dependence on color and stellar mass. We summarize our results and their implications in Section \ref{sec:conclusions}. 

Throughout this paper, we assume a flat $\Lambda$CDM cosmology as in \citet{Anderson:2012}, with $\Omega_\mathrm{m}=0.274$, $h=0.7$, $\Omega_\mathrm{b} h^2=0.0224$, $n_\mathrm{s}=0.95$ and $\sigma_8=0.8$. All comoving distances are quoted assuming $H_0=70\ \mathrm{km\ s^{-1}\ Mpc^{-1}}$.

\section{Methods} \label{sec:methods}

\subsection{Data}

\begin{figure*}[h!]
\epsscale{1.1}
\plottwo{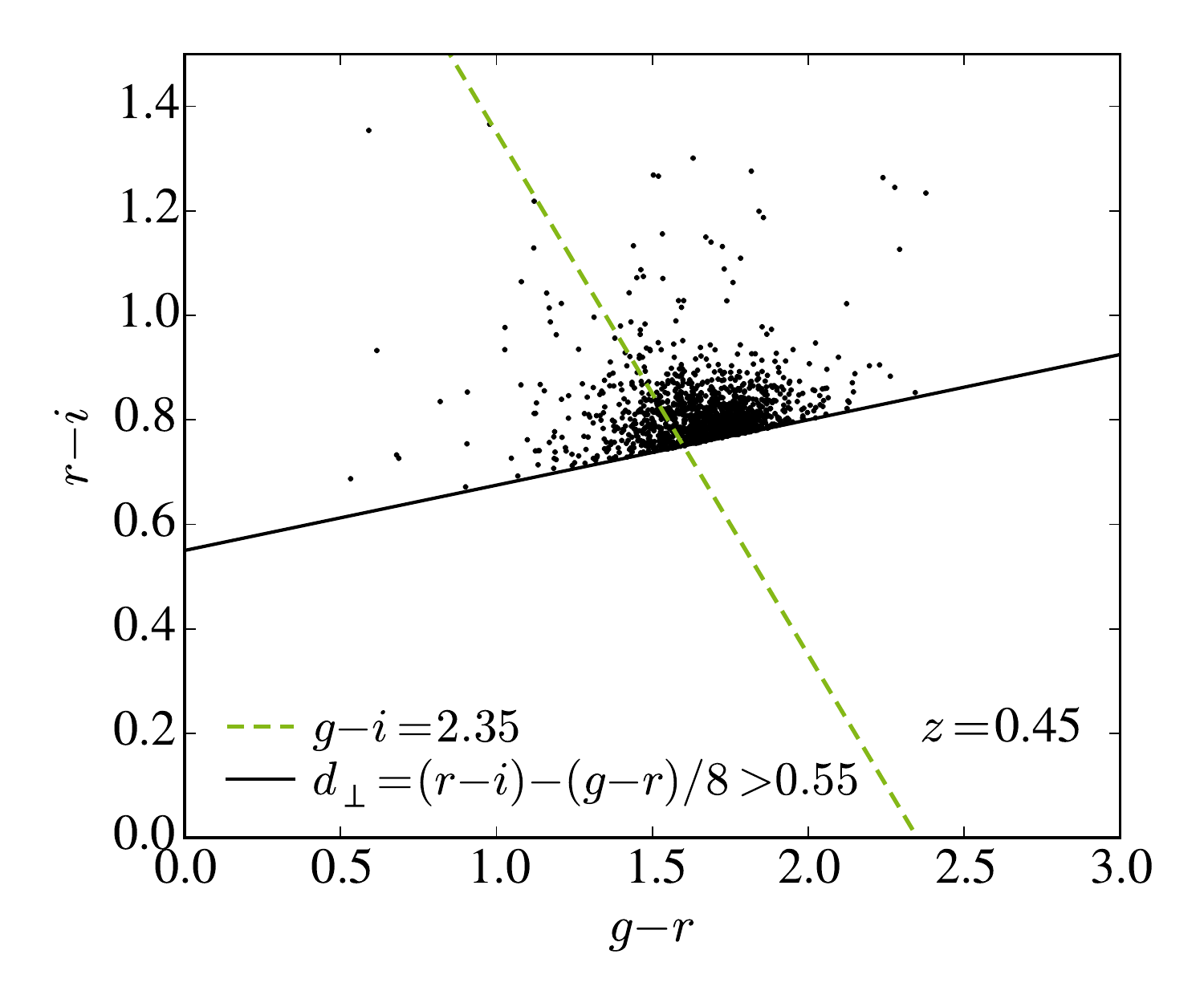}{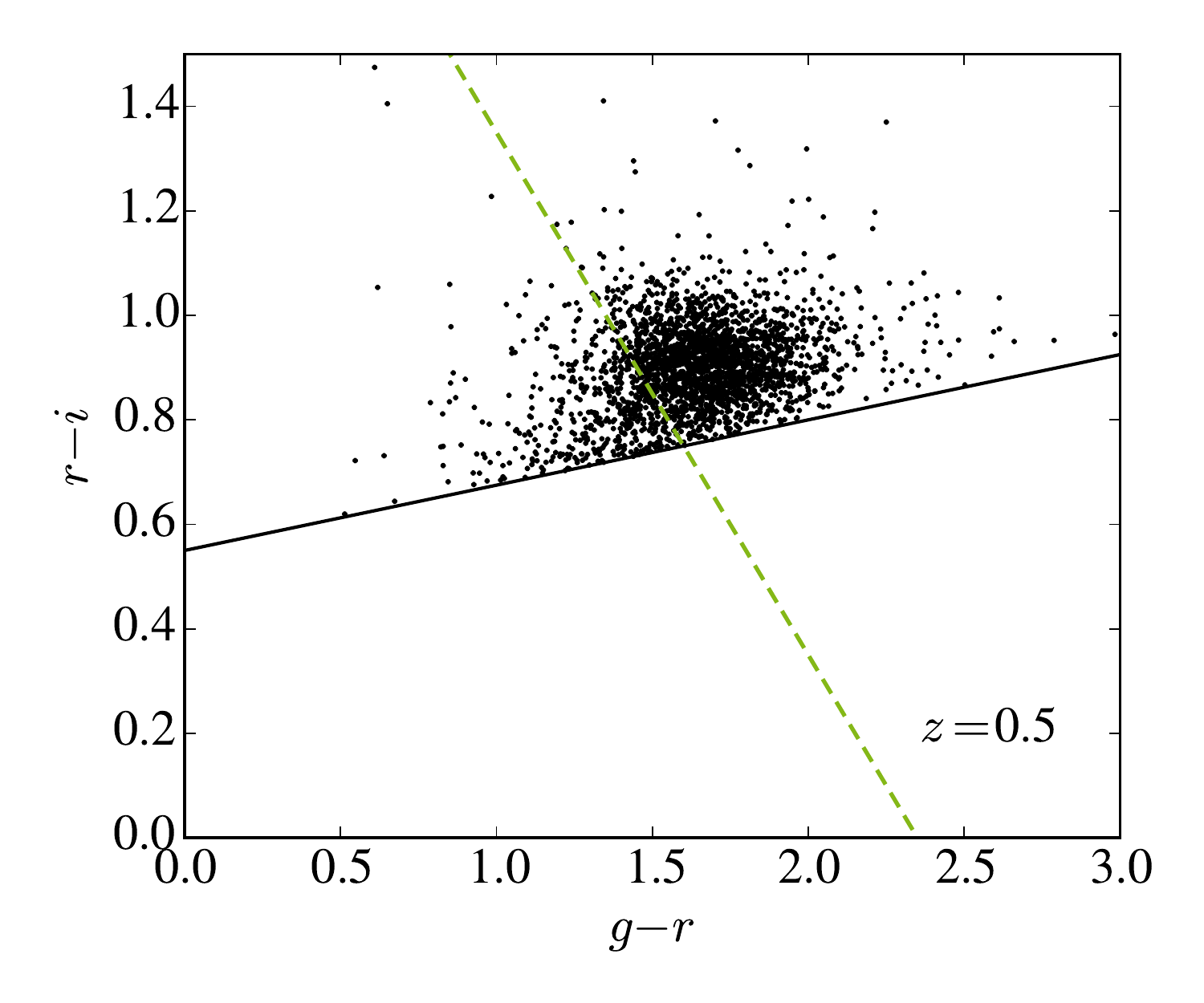}
\plottwo{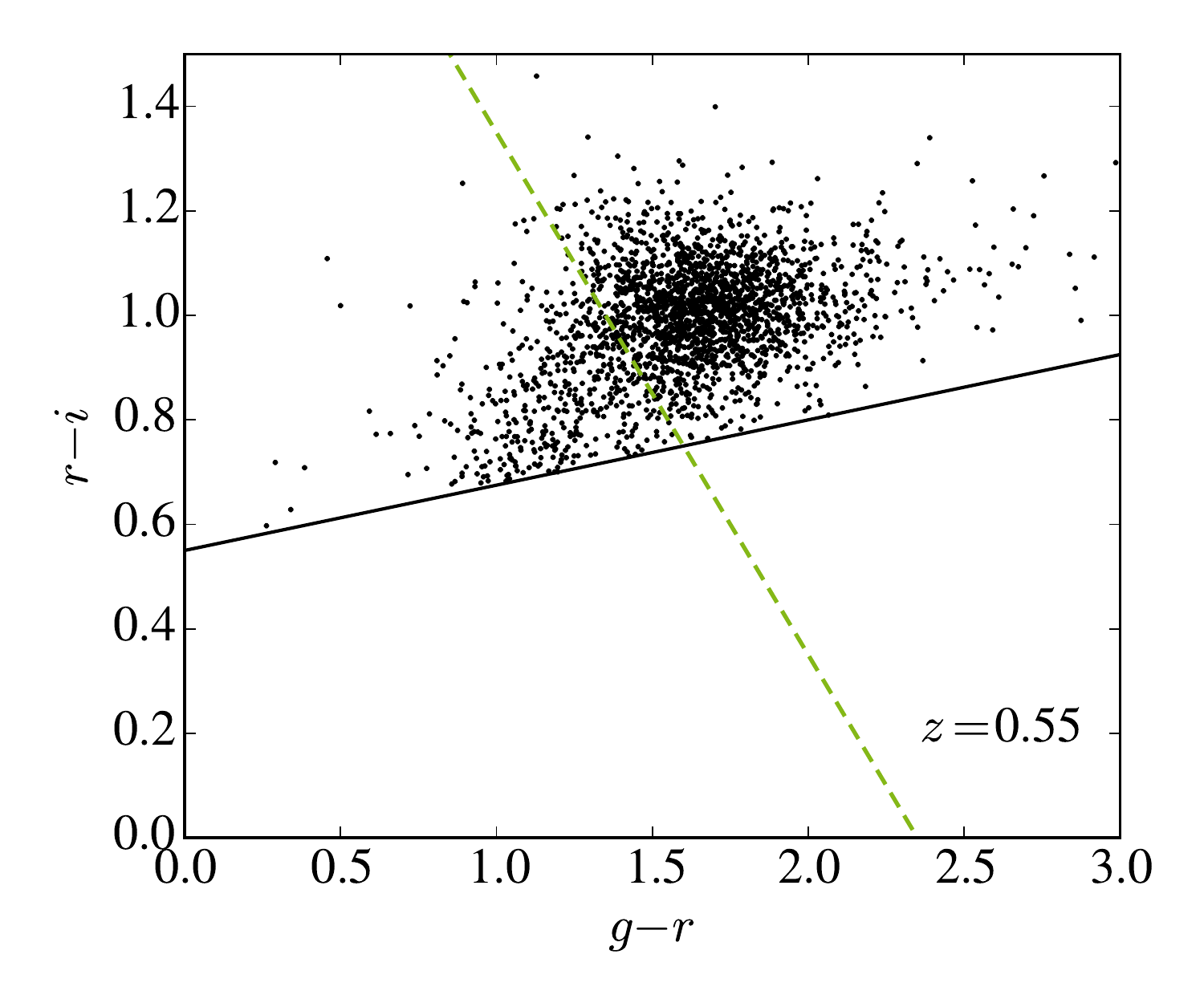}{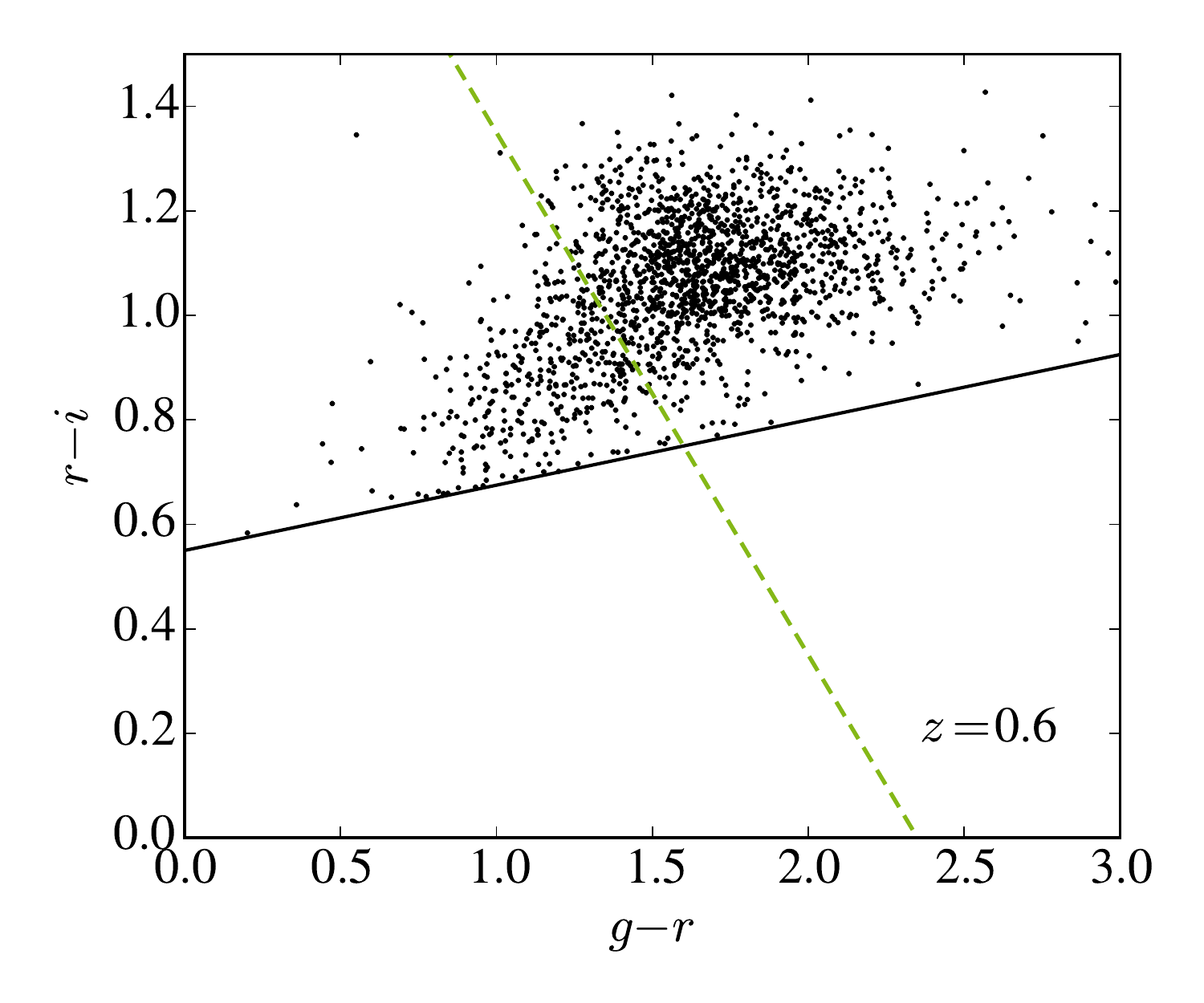}
\plottwo{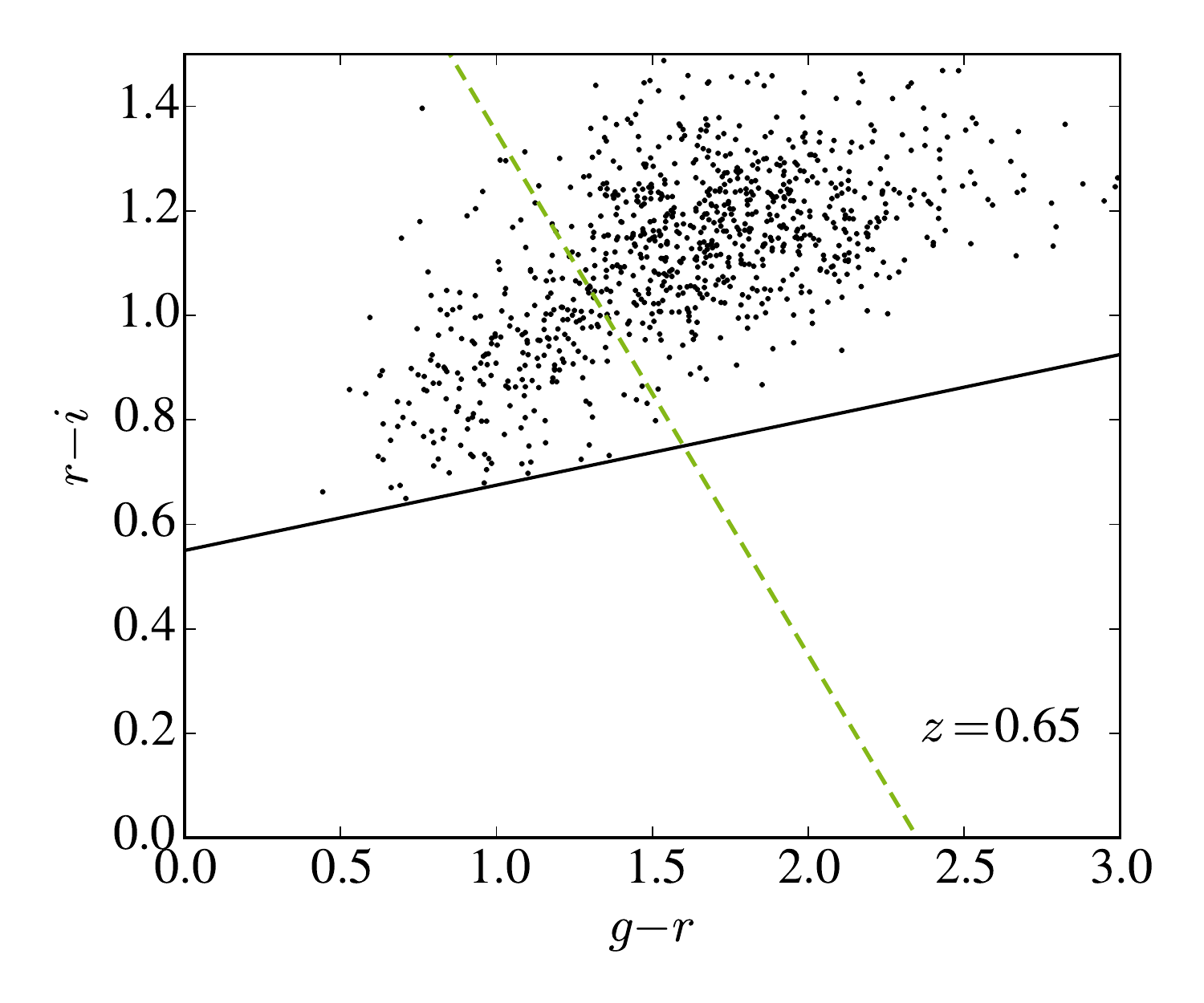}{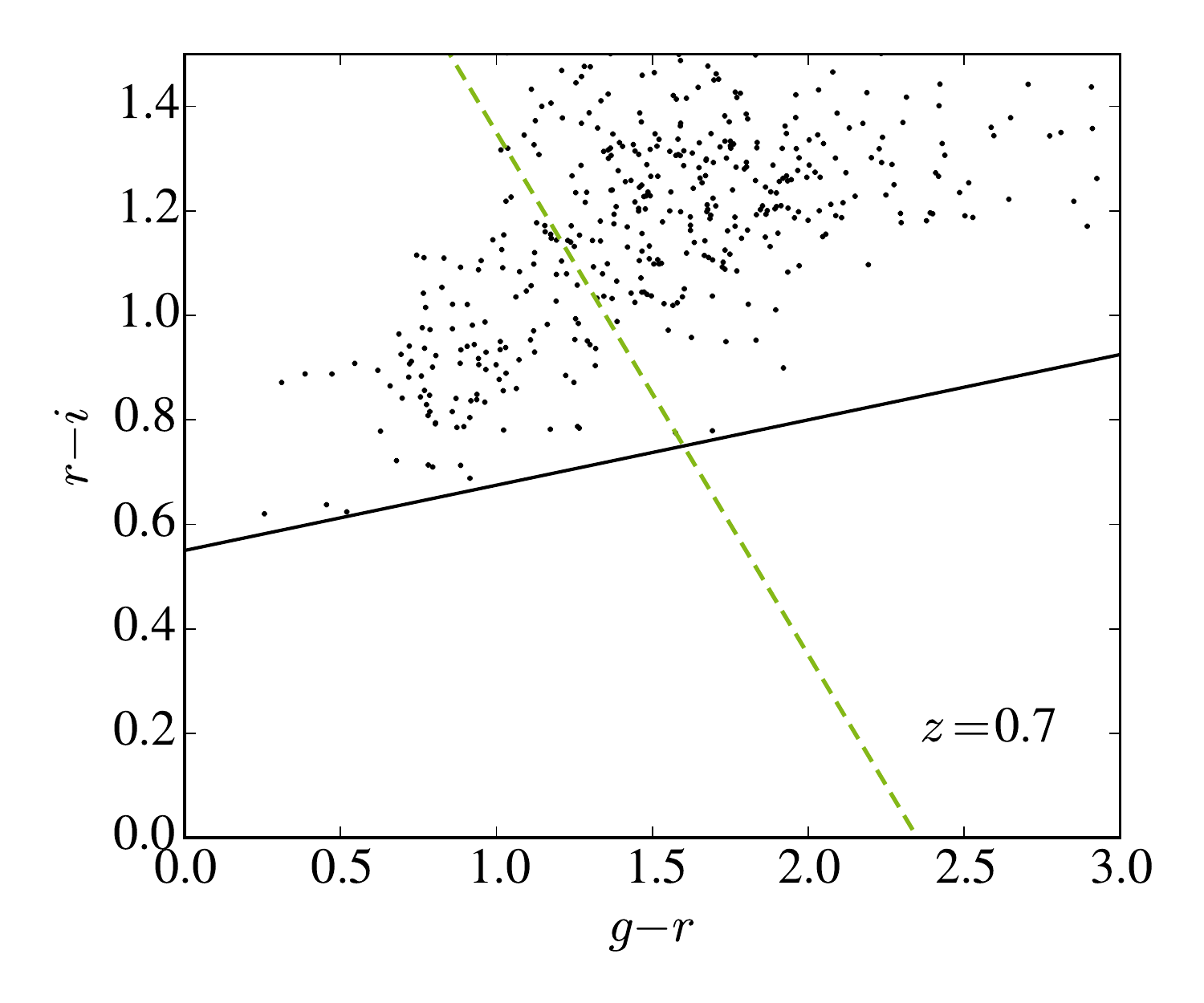}
\caption{Redshift progression of spectroscopic galaxies in $r - i$ vs. $g - r$. The $d_{\perp}$ cut is shown in solid black and the divide between blue and red galaxies in dashed green. At $0.6 < z < 0.65$, we have good populations of red and blue galaxies, with neither being cut off.}
\label{progression}
\end{figure*}

The SDSS survey \citep{York:2000} is a large-scale imaging and spectroscopic redshift survey, using a 2.5 m wide-angle optical telescope at Apache Point Observatory in New Mexico, USA \citep[for more details, see][]{Gunn:2006}. SDSS-III \citep{Eisenstein:2011} is the third phase of this project, and it includes BOSS \citep{Dawson:2013} among its four main surveys. SDSS uses a specially designed camera \citep{Gunn:1998} with five filters \citep[$u$, $g$, $r$, $i$, and $z$, with magnitude limits of 22.0, 22.2, 22.2, 21.3, and 20.5 respectively;][]{Fukugita:1996, Smith:2002, Doi:2010}; we use the $g$ and $i$ filters to assign galaxy color. Spectra are obtained using the double-armed BOSS spectrograph \citep{Smee:2013} and redshifts are determined as described in \citet{Bolton:2012}. BOSS covers over 10,000 square degrees of sky, obtaining the redshifts of 1.5 million luminous galaxies with $z < 0.8$. We use imaging catalogs from the full BOSS footprint as presented in DR9 \citep{Ahn:2012}, while we use spectroscopy from 

\begin{deluxetable}{lll}
\tablecaption{Properties of Spectroscopic and Imaging Data Sets\label{basic}}
\tablehead{
\colhead{} & \colhead{Spectroscopic} & \colhead{Imaging}}
\startdata
Apparent Magnitude & $< 19.9$ & $19.9 - 20.9$  \\ 
Number Density $[h^3 $Mpc$^{-3}]$ & $1.2\times 10^{-4}$ & $10^{-3}$ \\
Number of Galaxies & 66,657 & 6625,645 \\
Area [deg$^2$] & 6612 & 9432  \\
\enddata
\end{deluxetable}

\noindent the CMASS large-scale structure catalog of DR10 \citep{Ross:2012, Ahn:2014, Anderson:2014, Reid:2016}.

\subsubsection{Spectroscopic Data Set}
We use the CMASS sample to focus on the higher redshift galaxy sample of BOSS. The main color cut of the CMASS selection is $d_{\perp} = (r -i) - (g -r)/8 > 0.55$ \citep{Reid:2016}. We choose $g - i = 2.35$ as the cutoff between blue and red spectroscopic galaxies (blue being $g - i < 2.35$, red being $g - i > 2.35$). Observing the behavior of spectroscopic galaxies on an $r - i$ versus $g - r$ plot with increasing redshift, we see that this is a reasonable choice. This progression is shown in Figure \ref{progression}, with the $g-i=2.35$ line in dashed green and the $d_{\perp} >0.55$ cut in solid black. After $z=0.45$, galaxies tend to move parallel to the $g - i$ line with increasing redshift, which shows that our red and blue populations do not drift into one another. At lower redshifts, the sample (particularly in the blue) is partially cut off, but in our selected redshift range of $0.6 < z < 0.65$, there is a good population of red and blue galaxies. In calculating color-band fluxes, we correct for reddening due to Galactic extinction \citep[via the][dust maps]{Schlegel:1998}.

\begin{figure}
\epsscale{1.23}
\plotone{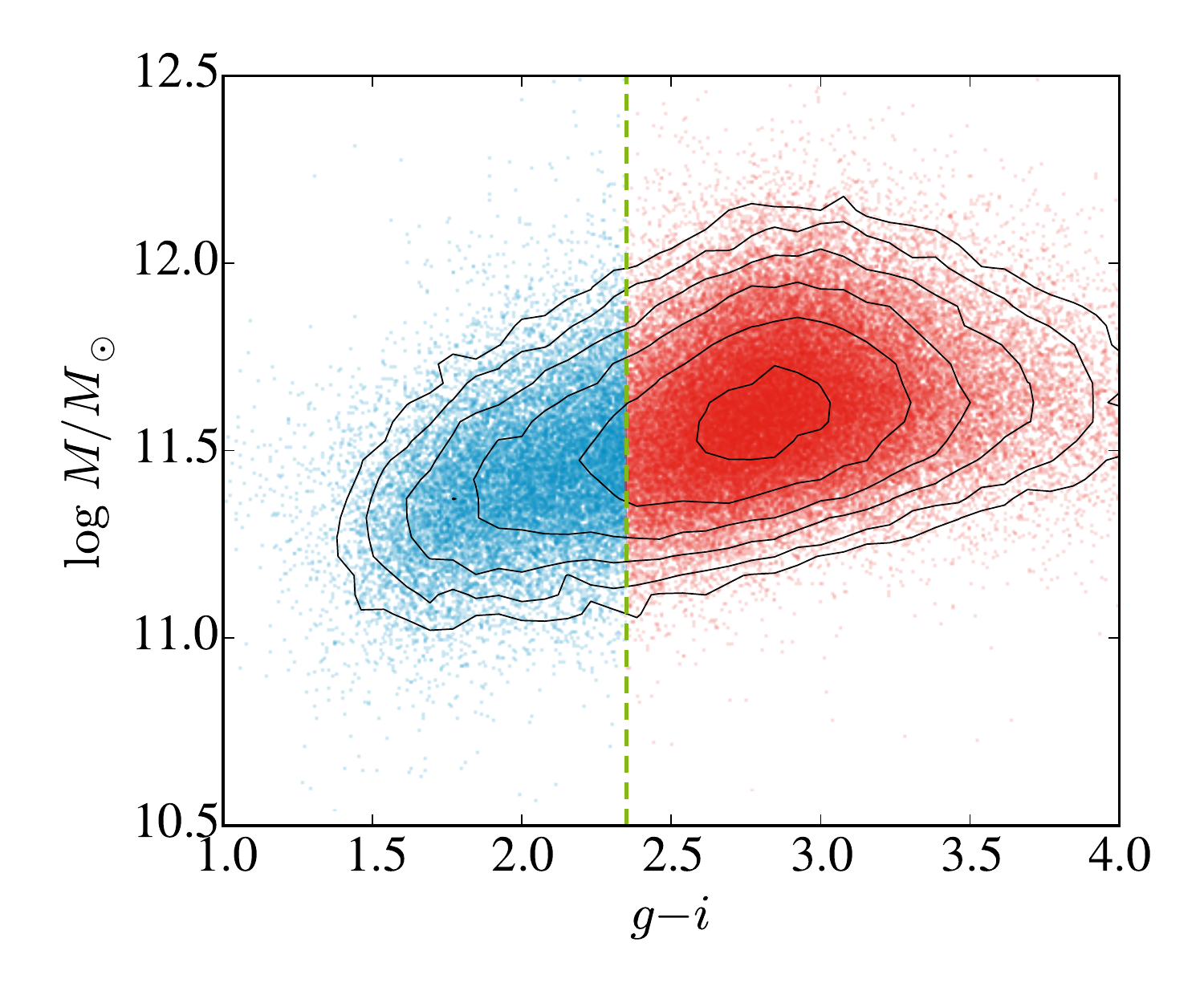}
\caption{Mass distribution vs. $g - i$ color for spectroscopic data set at $0.6 < z < 0.65$. Blue galaxies are on the left and red on the right of the dashed green line. Contours are spaced by a factor of 2.}
\label{mass_color}
\end{figure}

Our stellar masses are derived in \citet{Chen:2012}. These are the Wisconsin stellar masses, using templates from \citet{Bruzual:2003}. BOSS spectra are fit to a set of principal components derived from a large set of model spectra. From the mass-to-luminosity ratio of the best-fit model, the SDSS $i$ band c model magnitude is converted to stellar mass. The mass distribution of the spectroscopic data set is shown in Figure \ref{mass_color}, where it is plotted against color in $g - i$. Mass is measured as the logarithm of the galaxy stellar mass in solar units. Our blue galaxies are on average less massive than our red galaxies. We choose to isolate mass in bins of 0.25, which gives us three reliable mass bins in blue (from 11 to 11.75) and four in red (from 11.25 to 12.25). Two of these are overlapping, allowing us to compare clustering to color in fixed mass bins.

The spectroscopic data set has high luminosity ($i < 19.9$) and low number density \citep[$n= 1.2 \times 10^{-4}$ $h^3 \text{Mpc}^{-3}$ comoving, at our redshift;][]{Anderson:2012}. After redshift selection we are left with 66,657 spectroscopic galaxies, 14,480 of which are blue and 52,177 of which are red (see Table \ref{basic}). This sample is not numerous enough for precise auto-correlation functions on small scales; we therefore turn to cross-correlations between this and an imaging sample. By including the fainter imaging sample, we increase the number density of environment galaxies by a factor of ten. Conducting a cross-correlation analysis allows us to measure clustering in smaller radius annuli and to maintain low errors. 

\subsubsection{Photometric Data Set}

Our redshift selection of $0.6 < z < 0.65$ is easy to make on the spectroscopic galaxies as their redshifts are known; this is not the case for the imaging galaxies. Fortunately, we are only concerned with the excess correlation of imaging galaxies, which is unaffected by foreground or background galaxies. Any fluctuations in these foreground/background imaging galaxies are nearly uncorrelated with signals we are looking for in our redshift range. Small correlations can result from magnification from gravitational lensing or from consistent errors between the samples in the photometric calibration or dust extinction corrections; however, we expect these effects to be well below the strong small-scale intrinsic clustering at these redshifts \citep{Menard:2008}.

In practice, the background imaging galaxies are not very significant, simply because their faintness prevents high-$z$ galaxies from showing up on the survey. The imaging data set will contain very few galaxies at $z > 0.9$. On the other hand, $z < 0.45$ galaxies are eliminated by the $d_{\perp}$ cut. Here we require $d_{\perp} > 0.6$. This cut highly favors $z > 0.45$ galaxies, as the $4000~\textrm{\AA}$ break is in the $r$ band, making the $r-i$ color redder with increasing redshift. We have a tighter requirement on $d_{\perp}$ for imaging galaxies for two reasons: (1) the scatter in observed colors is larger for faint objects, making it more likely that low-redshift objects scatter into the cut, and (2) we have selected a $z>0.6$ spectroscopic sample, not just $z>0.45$. It is useful to reduce the number of high- and low-redshift imaging galaxies because while they do not contribute to our correlation signal, they do contribute to the noise (by increasing the number of objects in our radius bins, we increase the variance in the counts). We are left with 6625,645 imaging galaxies with magnitudes of $19.9 < i < 20.9$ and an average number density of $n = 10^{-3}$ $h^{3}\text{Mpc}^{-3}$ (see Table \ref{basic} for a summary).

While we expect the redshift distribution of the imaging data set to be similar to that of the spectroscopic \citep[shown in Figure 2 of][]{Anderson:2012}, we do not know it exactly. This is unimportant for our calculation, as we are concerned with differential comparisons between spectroscopic galaxies, making only the ratio of their environments important. Any uncorrelated addition to the density of imaging galaxies around each spectroscopic galaxy will not impact excess correlation---the parameter we are after.

The survey accounts for---and minimizes the potential systematic error from---variations due to Galactic extinction, seeing, stellar density, sky background, airmass, and photometric offset. This includes, for example, a mask for the presence of stars of comparable magnitudes to our galaxies, without removing any galaxies from the sample. We use a map of the masked regions of the imaging data set defined in \citet{Ross:2011}; the strategy is described in the next section.

\subsection{Strategy}

\begin{figure}[t!]
\epsscale{0.665}
\plotone{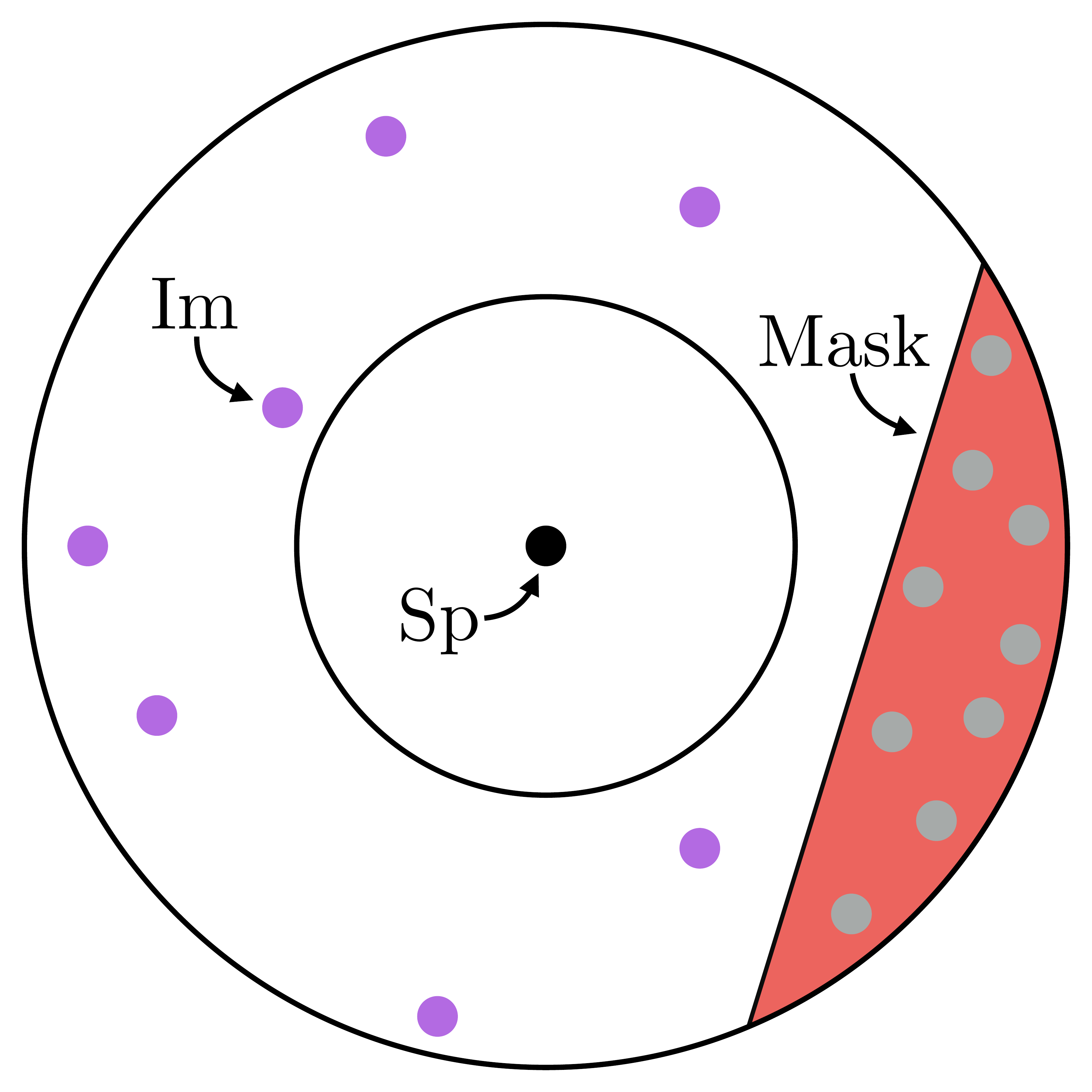}
\caption{Cartoon of counting setup. The inner and outer rings represent the bounds of a particular radius annulus, with the masked region in red (enlarged here for clarity). The spectroscopic galaxy is in black, the imaging galaxies in purple, and the masked points in gray.}
\label{cartoon}
\end{figure}

Our goal is to determine how the clustering of the spectroscopic galaxies changes with their color and stellar mass. First, we define color according to the difference in magnitude in the $g$ and $i$ bands: we denote $g-i < 2.35$ as blue and $g-i > 2.35$ as red. We then count the number of imaging galaxies around each spectroscopic galaxy, comparing between the blue and red selections. We do this for different angular bins in order to assess how the clustering dependence on color changes with radius. The annuli, in terms of proper radii from each spectroscopic galaxy, are: 125--250, 250--500, 500--1000 kpc, 1--2, and 2--4 Mpc. Knowing that clusters are predominantly composed of red galaxies, we might expect these to have denser environments than blue galaxies. We perform this same analysis in different mass bins, keeping color fixed. We also compare clustering versus color in fixed mass bins.

Our first problem is that the survey is imperfect---many regions are masked due to foreground stars or poor imaging. We generate random points in the masked regions (meaning in the excluded regions, opposite of common practice). We then count the number of points that appear in a radial bin so as to make an estimate of the fraction of the bin than is masked out. We embed the survey in a generous oversized bounding box, such that the masked area is 6108 deg$^2$, then generate 17,767,741 points in the masked area. For each spectroscopic galaxy, the picture looks something like Figure \ref{cartoon}, where we count the number of imaging galaxies within a specified radius annulus of the galaxy, accounting for the masked regions (enlarged here for clarity).

Accounting for the mask, the average number of imaging galaxies around each spectroscopic galaxy should scale as $N_{\text{Sp},j} \propto (1 + w)A_\text{total}(1-f_\text{masked}) = (1 + w)(A_\text{total} - A_\text{mask}$), where $A_\text{total}$ is the total area between the inner and outer radii from the spectroscopic galaxy, $A_\text{mask}$ is the masked area in this region, and $f_\text{masked}=A_\text{mask}/A_\text{total}$. If this number of counts is uncorrelated with whatever selection $j$ we make on the spectroscopic galaxies, we expect the clustering amplitude $w(R)$ to be zero. The stronger the correlation of the clustering with our selection $j$, the larger $w(R)$ is. In this paper, we ask how $w(R)$ changes with color and stellar mass of the spectroscopic galaxies at different radii. We discuss the calculation of $w(R)$ more quantitatively below.

We define the average areal density of the imaging data set to be $n_\text{I} = N_\text{I}/A_\text{I}$, and the average areal density of the mask to be $n_\text{M} = N_\text{M}/A_\text{M}$. For our data sets, $n_\text{I} = 702$/deg$^2$ and $n_\text{M} = 2909$/deg$^2$. For a spectroscopic galaxy $j$ and a specific annulus between $\theta_\text{in}$ and $\theta_\text{out}$, we expect the following counts, $N$, of imaging galaxies:

\begin{deluxetable*}{llllll}
\tablecaption{Average Imaging Galaxy Counts $\left\langle N_{j}\right\rangle$ for Red and Blue Galaxies in Different Annuli\label{counts}}
\tablehead{
\colhead{$R_\text{inner}$ [kpc]\tablenotemark{a}} & \colhead{$R_\text{outer}$\tablenotemark{a}} & \colhead{$A$ [arcmin$^2$]\tablenotemark{b}} & \colhead{$f_\text{masked}$\tablenotemark{c}} & \colhead{$\left\langle N_{j} \right\rangle_\text{blue}$} & \colhead{$\left\langle N_{j} \right\rangle_\text{red}$}} 
\startdata
62.5 & 125 & 0.22 & 0.000 & 0.09 & 0.14 \\ 
125 & 250 & 0.86 & 0.012 & 0.27 & 0.36 \\ 
250 & 500 & 3.45 & 0.024 & 0.88 & 1.03 \\ 
500 & 1000 & 13.80 & 0.048 & 3.05 & 3.27 \\ 
1000 & 2000 & 55.20 & 0.077 & 11.20 & 11.51 \\ 
2000 & 4000 & 220.81 & 0.089 & 42.74 & 43.38 \\ 
\enddata
\tablenotetext{a}{The inner and outer radii of each annulus.}
\tablenotetext{b}{The total area of each annulus.}
\tablenotetext{c}{The fraction of this area that is masked.}
\end{deluxetable*}

\begin{eqnarray}
N_{j}&=&\left(1+w_j\right)n_\text{I}\ A_\text{total}(1 - f_\text{masked})\nonumber\\
&=&\left(1+w_j\right)n_\text{I} \left(A_\text{total} - A_\text{mask}\right)\nonumber\\
&=&\left(1+w_j\right)n_\text{I}\left(\pi(\theta_\text{out}^2-\theta_\text{in}^2)-N_{\text{M}, j}/n_\text{M}\right),\nonumber\\
&
\end{eqnarray}

\noindent where $N_{\text{M},j}$ is the number of random points in the masked region in the given annulus around the spectroscopic galaxy $j$. Averaging over the set of $j$ spectroscopic galaxies, we have

\begin{eqnarray}
\left\langle N_{j}\right\rangle &= \left(1+\left\langle w_j \right\rangle \right)n_\text{I}\left(\pi(\theta_\text{out}^2-\theta_\text{in}^2)-\left\langle N_{\text{M}, j}\right\rangle/n_\text{M}\right),\nonumber\\
& 
\end{eqnarray}

\noindent where $\left\langle x_i \right\rangle = \sum q_i x_i / \sum q_i$ with weights $q_i$ assigned as in Equation 18 of \citet{Anderson:2014} to correct for the effects of fiber collisions and redshift failures. From this we can easily solve for $w(R)$.

We calculate errors on $w(R)$ due to both Poisson variance (or shot noise) and sample variance. Shot noise is proportional to $\sqrt{N_\text{gal}}$, giving a signal-to-noise ratio proportional to $N_\text{gal}/\sqrt{N_\text{gal}} = \sqrt{N_\text{gal}}$. Sample variance derives from the fact that not all regions of the sky are identical, and so one patch (sample) of the sky will vary slightly from the next. We calculate sample variance by binning the sky into 100 non-overlapping areas and computing the variance in $w(R)$ between these areas. We calculate the jackknife errors, which include both Poisson variance and sample variance, as

\begin{eqnarray}
\sigma^2(R_i, R_j) = {N-1 \over N} \sum_{k=1}^N
&\left[ w_k(R_i) - \overline w_k(R_i) \right] \times \nonumber\\
&\left[ w_k(R_j) - \overline w_k(R_j) \right],
\end{eqnarray}

\noindent where $N=100$ regions, $w_k(R_i)$ is $w(R)$ calculated in the $i$th angular bin with region $k$ removed, and $\overline w_k(R_i)$ is the average of these $w_k(R_i)$ values. At smaller angular scales, our errors are dominated by Poisson variance, while sample variance dominates at the larger scales. Note that our errors account for the covariances between different angular scales.

\section{Results} \label{sec:results}
Table \ref{counts} shows the average number of imaging counts around each spectroscopic galaxy for different bins in radius and as a function of color. It also lists the corresponding areas and the fractions of these areas that are masked, which are consistently less than 10\%. We convert angles on the sky to transverse distances for each annulus using $z = 0.625$, the average redshift of our selection. The radii quoted are proper distances.

There are on average more imaging neighbors to red spectroscopic galaxies than to their blue counterparts. Table \ref{w} presents this information in terms of the clustering amplitude $w(R)$, along with the errors associated with its calculation. We present the jackknife errors as well as the Poisson errors (which are included in the former) for comparison. We include extra significant figures so that one can compare the two error estimates.

\begin{figure}[t]
\epsscale{1.23}
\plotone{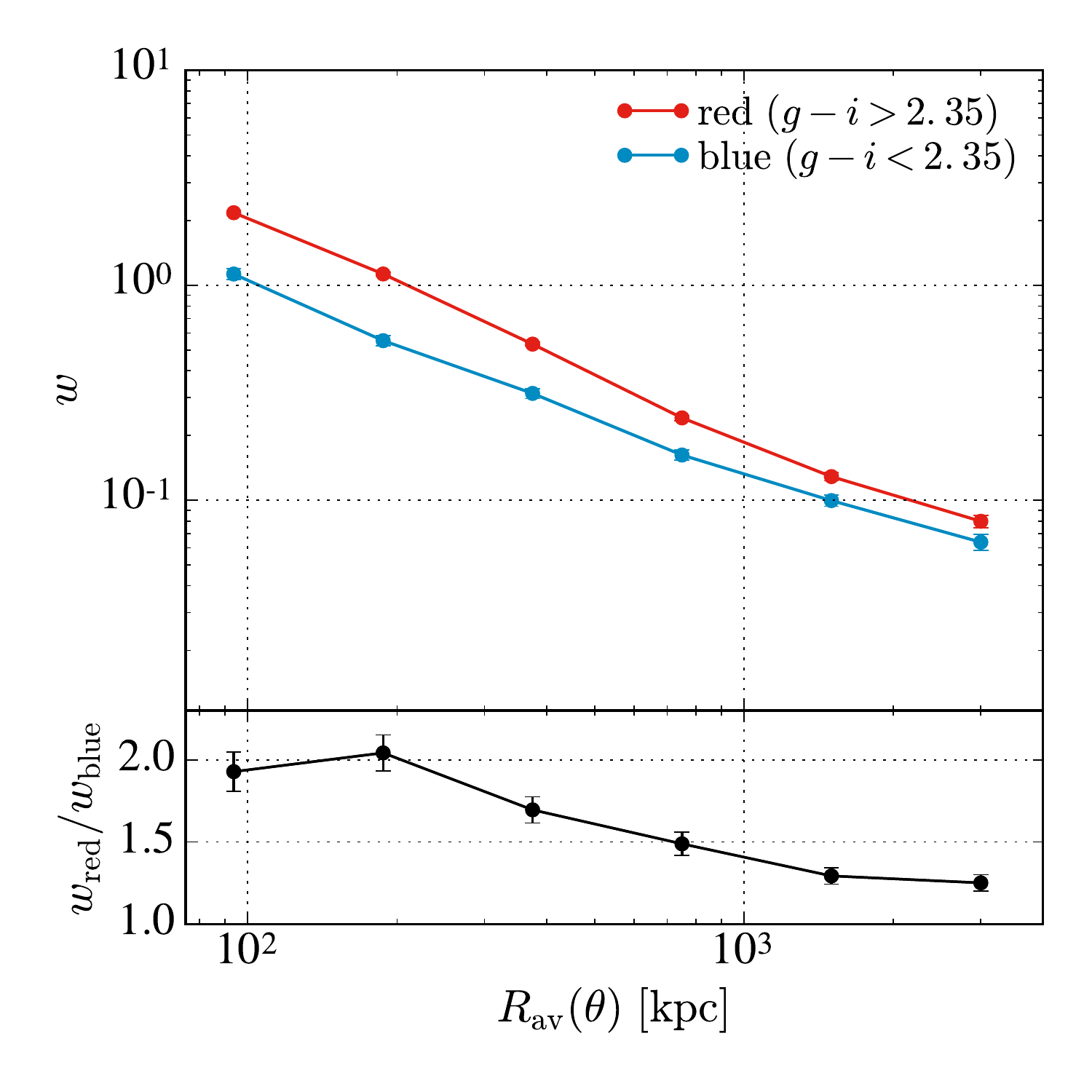}
\caption{Top panel: clustering amplitude $w(R)$ for red and blue galaxies in different annuli, using the full sample without mass cuts. Bottom panel: the ratio of these two amplitudes. The $x$ axis plots $R_\text{av}$ (in kpc), the average radius in each non-overlapping annulus.}
\label{w_v_r}
\end{figure}

We show the clustering for red and blue galaxies as a function of radius in Figure \ref{w_v_r}. $R_\text{av}$ denotes the average radius of a particular annulus. Note that the annuli are non-intersecting. We expect $w(R)$ to be greater for red galaxies than blue ones, as galaxy clusters are made predominantly of the former. $w(R)$ is much larger at smaller radii, indicating that the correlation between color and clustering is stronger at smaller scales as expected. With increasing $R$, $w(R)$ approaches zero. It

\begin{deluxetable*}{llllll}
\tablecaption{Clustering Amplitudes $w(R)$ and Errors for Red and Blue Galaxies in Different Angular Annuli\label{w}}
\tablehead{
\colhead{$R_\text{av}$ [kpc]\tablenotemark{a}} & \colhead{$w_\text{blue}\pm \sigma_{w,\text{blue}}$\tablenotemark{b}} & \colhead{$\sigma_{w,\text{blue,Pois}}$\tablenotemark{c}} & \colhead{$w_\text{red}\pm \sigma_{w,\text{red}}$} & \colhead{$\sigma_{w,\text{red,Pois}}$} & \colhead{$w_\text{red}/w_\text{blue}$\tablenotemark{d}}} 
\startdata
93.75 & $1.1282 \pm 0.0637$ & $\pm 0.0602$ & $2.1755 \pm 0.0481$ & $\pm 0.0382$ & $1.93 \pm 0.12$   \\ 
187.5 & $0.5526 \pm 0.0294$ & $\pm 0.0263$ & $1.1287 \pm 0.0201$ & $\pm 0.0164$ & $2.04 \pm 0.11$   \\ 
375 & $0.3140 \pm 0.0168$ & $\pm 0.0127$ & $0.5324 \pm 0.0104$ & $\pm 0.0074$ & $1.70 \pm 0.08$   \\ 
750& $0.1623 \pm 0.0090$ & $\pm 0.0066$ & $0.2416 \pm 0.0069$ & $\pm 0.0036$ & $1.49 \pm 0.07$     \\ 
1500 & $0.0996 \pm 0.0060$ & $\pm 0.0037$ & $0.1288 \pm 0.0057$ & $\pm 0.0020$ & $1.29 \pm 0.05$   \\
3000 & $0.0638 \pm 0.0055$ & $\pm 0.0023$ & $0.0798 \pm 0.0052$ & $\pm 0.0012$ & $1.25 \pm 0.05$    \\ 
\enddata
\tablenotetext{a}{The average radius of each annulus.}
\tablenotetext{b}{Jackknife errors.}
\tablenotetext{c}{Poisson errors (included in jackknife errors).}
\tablenotetext{d}{Errors in $w_\text{red}/w_\text{blue}$ are derived through the jackknife method with the ratios in individual jackknife samples.}
\tablecomments{The ratio $w_\text{red}/w_\text{blue}$ increases with decreasing $R$.}
\vspace{-0.5cm}
\end{deluxetable*}

\begin{figure}[t]
\epsscale{1.23}
\plotone{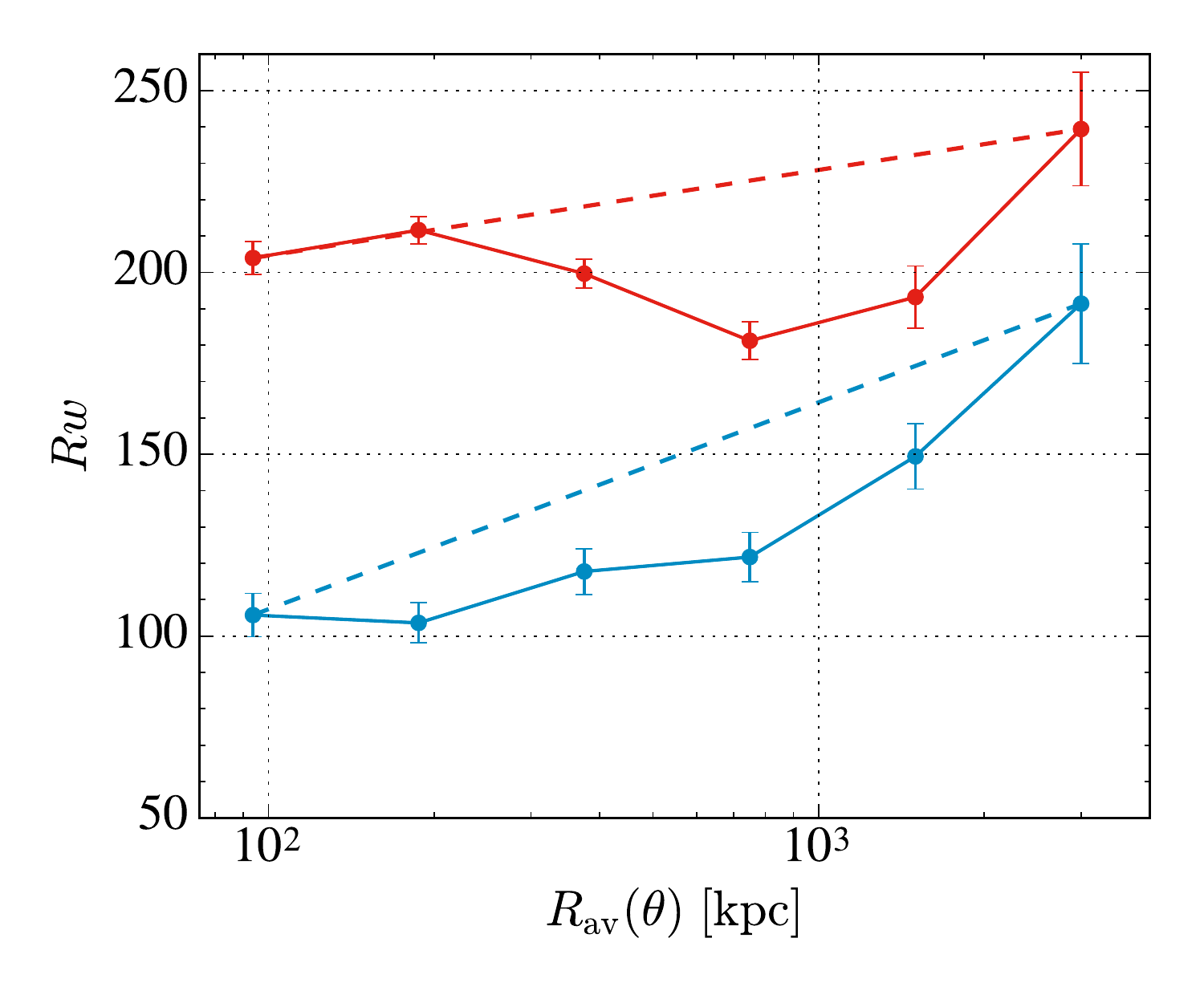}
\caption{$Rw$ vs. $R$ for red ($g-i>2.35$) and blue ($g-i<2.35$) galaxies in different annuli, using the full sample without mass cuts. $w(R)$ and $R$ do not obey a simple power-law relation, with a pronounced dip around 1 Mpc.}
\label{rw_v_r}
\end{figure}

\noindent is also useful to observe the ratio of $w_\text{red}$ to $w_\text{blue}$ (this is more interesting than simply the difference in magnitude between the two). This ratio is largest at small $R$, and it decreases with increasing $R$. As we increase the scale to greater than that of a galaxy cluster, the clustering dependence on color should weaken. $w_\text{red}/w_\text{blue}$ decreases as $R$ increases.

\begin{figure*}[t]
\epsscale{1.17}
\plottwo{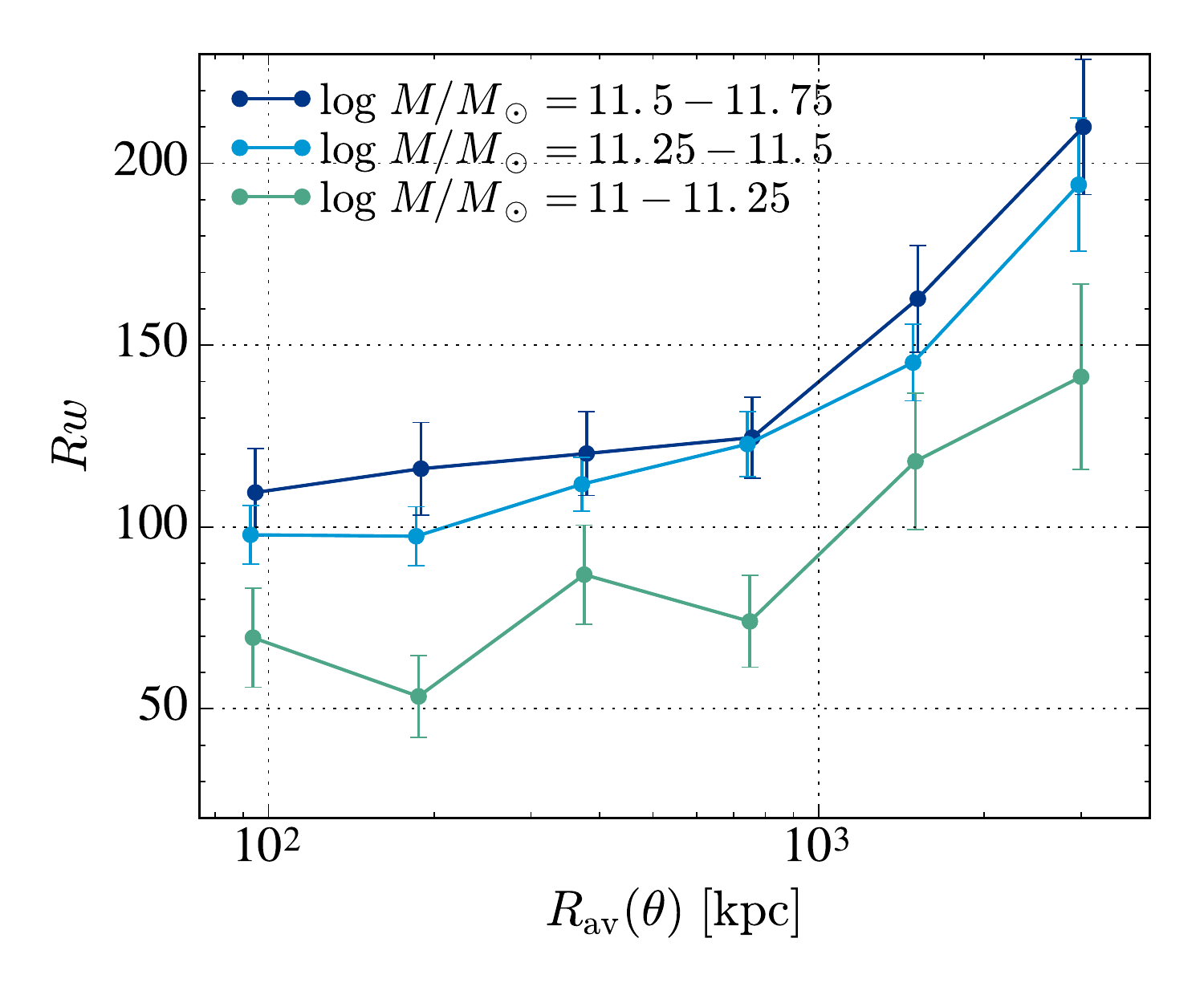}{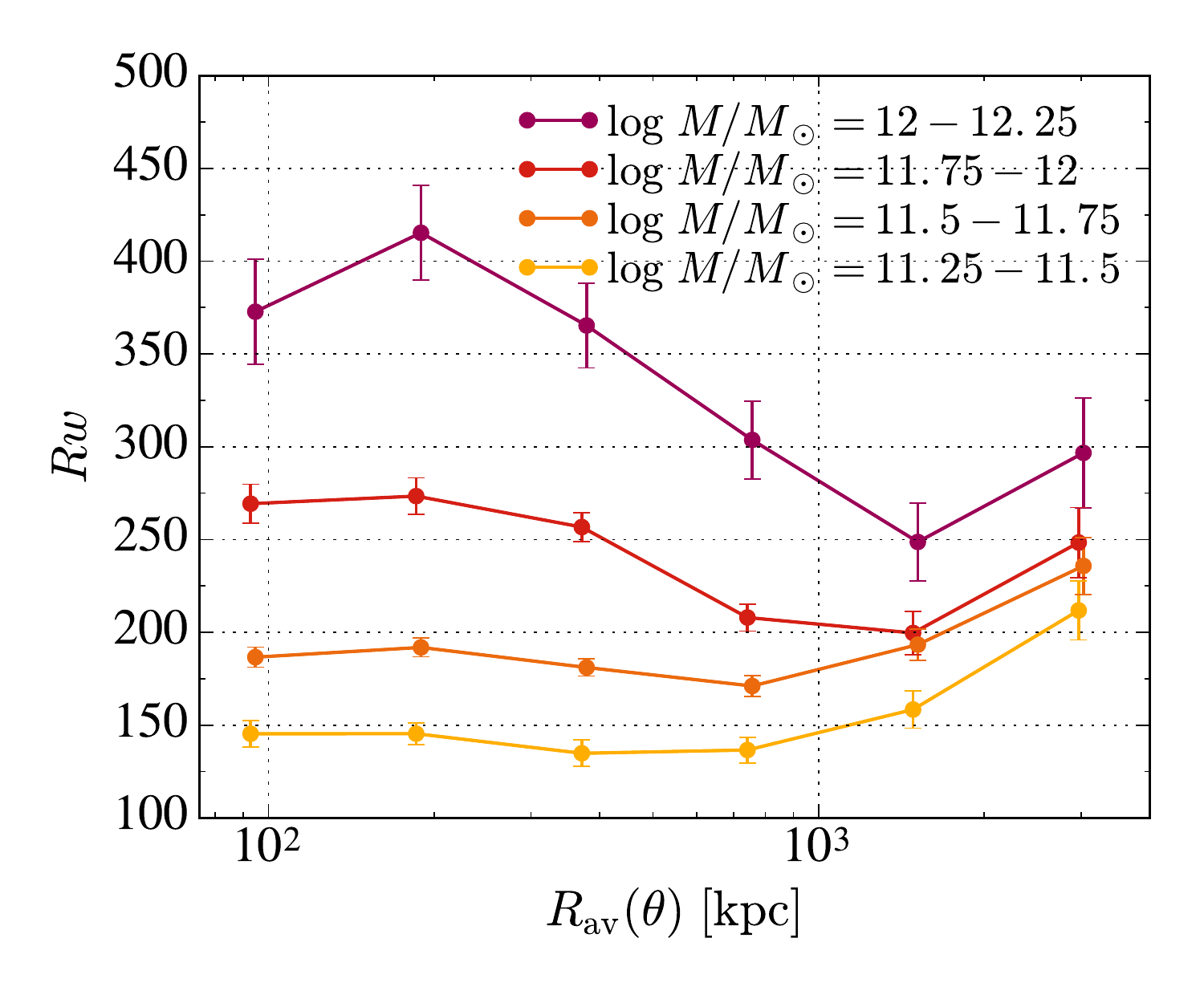}
\caption{Left panel: $Rw$ vs. $R$ for blue galaxies in different mass bins. Clustering amplitude increases with increasing mass, but there is not much difference between the highest two mass bins. Right panel: $Rw$ vs. $R$ for red galaxies in different mass bins. Clustering amplitude increases more than twofold from lowest to highest mass at small scales.}
\label{fixed_color}
\end{figure*}

\begin{figure*}[t]
\epsscale{1.17}
\plottwo{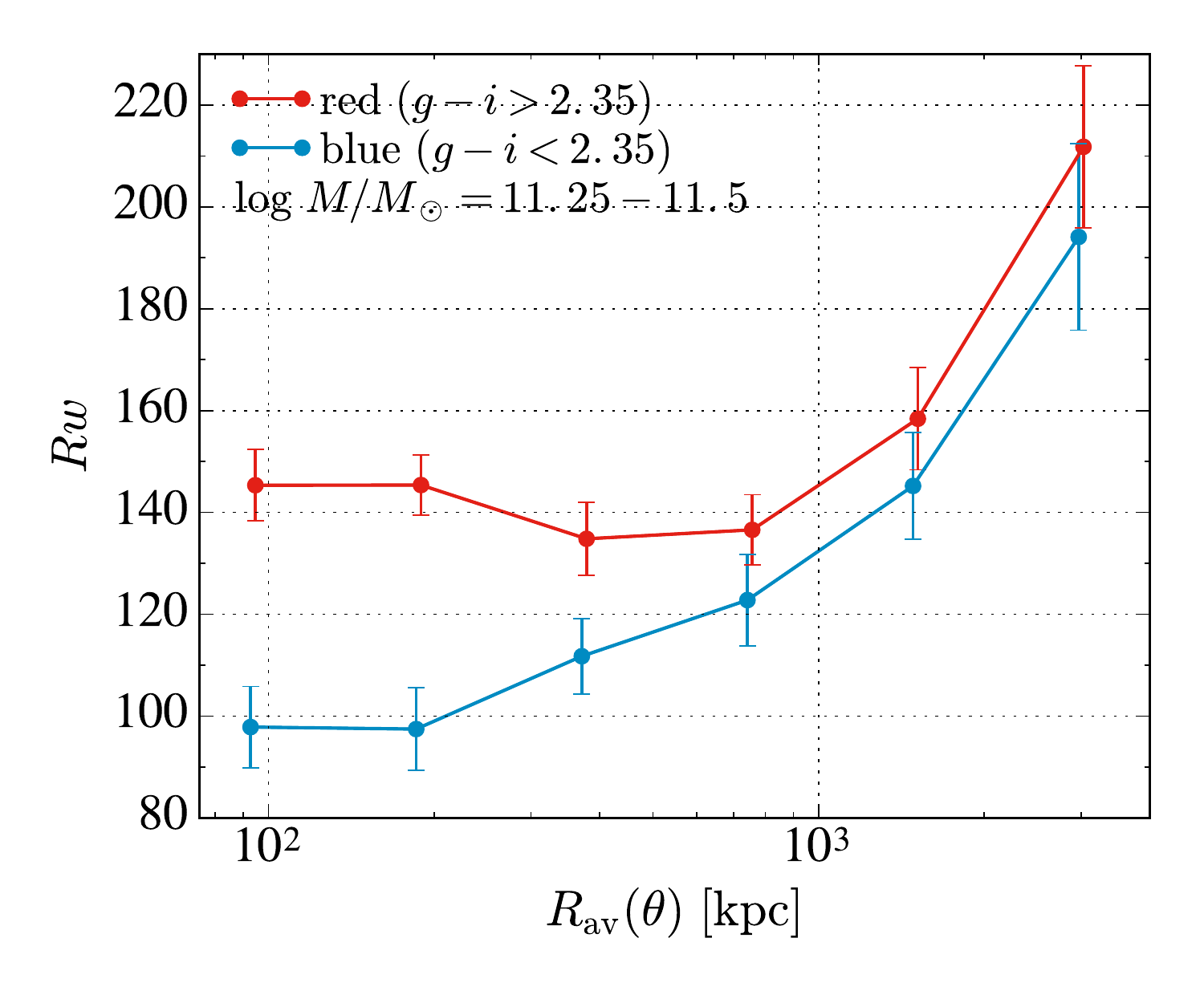}{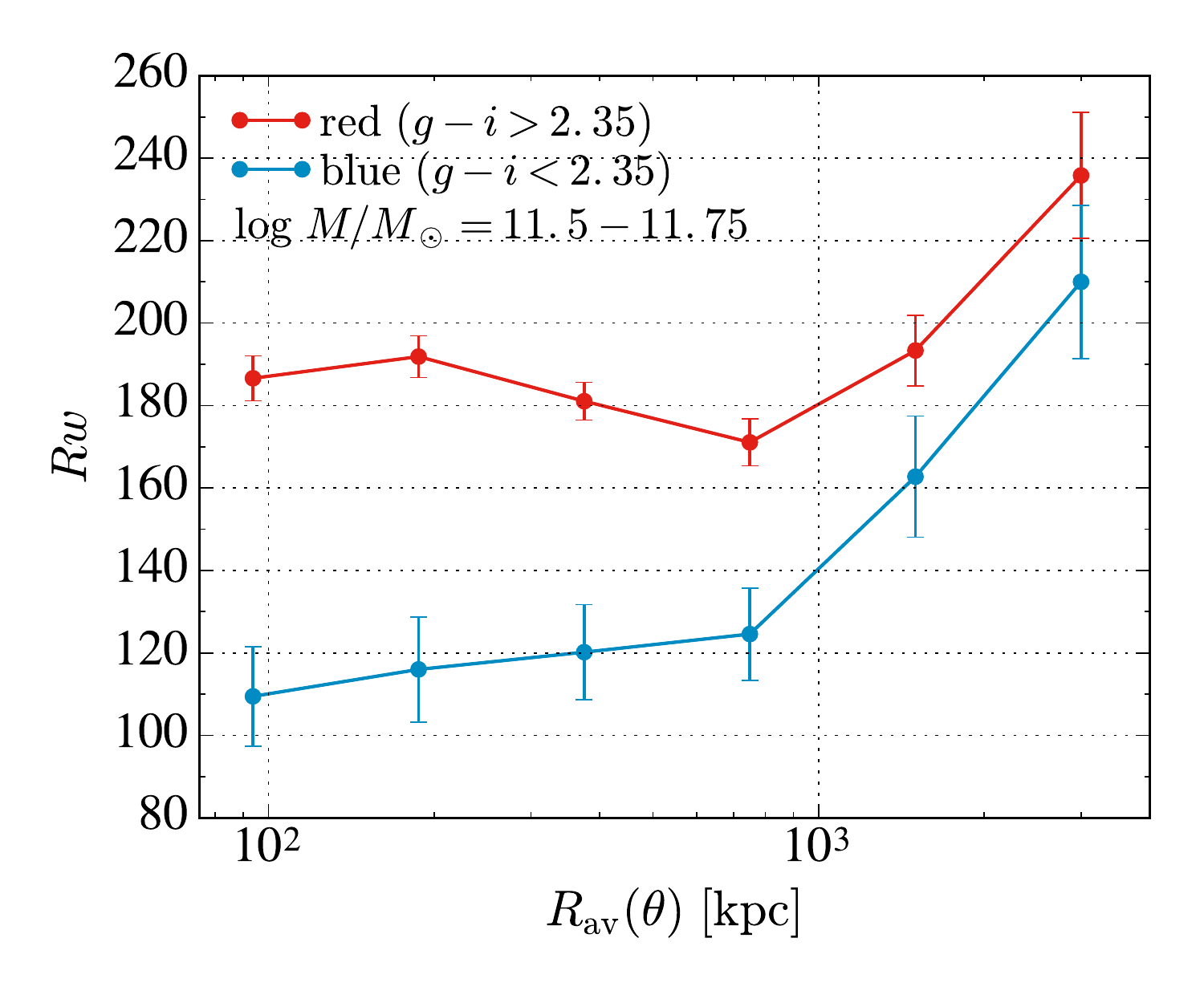}
\caption{$Rw$ vs. $R$ for red and blue galaxies with $\log(M/M_\odot)$ between 11.25 to 11.50 in the left panel and 11.50 to 11.75 in the right panel. Holding mass fixed, red galaxies show a strong enhancement in clustering amplitude at sub-Mpc scales.}
\label{fixed_mass}
\end{figure*}

To focus on more subtle trends, we plot $Rw$ as a function of $R$ in Figure \ref{rw_v_r}. We see that $w(R)$ and $R$ do not have a simple power-law relationship---if this was the case, the curve would be straight and monotonic, as is suggested by the dashed lines connecting the first and last points. Around $R = 1$ Mpc, there is a pronounced dip in the curve, especially for red galaxies. This is interpreted in the HOD formalism as a transition to a two-halo regime \citep{Zehavi:2005}. It makes sense that this should happen around 1 Mpc, the scale of a galaxy cluster.

We next investigate how galaxy clustering depends upon stellar mass. We do this by measuring the clustering amplitude in different mass bins for fixed color. Figure \ref{fixed_color} shows this relationship for blue galaxies in the left panel and red galaxies in the right panel. We label our mass bins as the logarithm of the stellar mass in solar units. For blue galaxies, we find a positive clustering amplitude--stellar mass relation between the 11-11.25 and 11.25-11.5 mass bins but not much differentiation between the 11.25-11.5 and 11.5-11.75 mass bins. For red galaxies, the mass variation on larger scales is mild (but still of order 30\%), indicating only a moderate change in the large-scale bias. The small-scale clustering amplitude is much stronger for the massive galaxies, where there is more than a twofold increase in amplitude from the 11.25-11.5 to 12-12.25 mass bins. In both the blue and red samples, there is the smallest difference in amplitude between the 11.25-11.5 and 11.5-11.75 mass bins. If we accept that higher amplitude suggests higher dark matter halo mass, our findings indicate that higher stellar mass red galaxies are in increasingly large dark matter halos. This is not statistically true for blue galaxies between all mass bins. \citet{Skibba:2015}, consistent with previous observations and theoretical predictions, find a strong correlation between central galaxy stellar mass and dark matter halo mass. Their highest stellar mass bin has $\left\langle{\log(M/M_\odot)}\right\rangle=11.20$; we extend the relationship to bins with higher average stellar masses.

Another interesting question is whether the clustering amplitude--color relation we found above holds for fixed mass. Figure \ref{fixed_mass} shows $Rw$ versus $R$ plots for galaxies with $\log(M/M_\odot)$ bins of 11.25-11.5 in the left panel and 11.5-11.75 in the right panel. There is a clear difference in clustering amplitude between red and blue galaxies in the same mass bin, especially at smaller scales. This shows that clustering amplitude is not solely dependent on stellar mass. \citet{Mandelbaum:2016} use galaxy--galaxy lensing to study the dark matter halos surrounding a sample of bright SDSS galaxies. They find that quiescent central galaxies have halos that are at least twice as massive as those of star-forming galaxies of the same stellar mass, and that this effect exceeds $3\sigma$ for $\log(M/M_\odot)>10.7$, i.e., for our mass bins. While their sample extends only to $\log(M/M_\odot)<11.6$, there is enough overlap to confirm that this bimodality at fixed mass also exists in galaxy clustering. Besides the halo mass difference, on small scales there may also be contributions from the difference in the satellite fraction of red and blue CMASS galaxies. \citet{Guo:2014} require a fraction of the galaxies to be satellites in massive halos in order to explain strong small-scale clustering.

\citet{Guo:2013, Guo:2014} conducted similar studies to ours, using the same spectroscopic data set (CMASS) and redshift range, but using auto-correlation rather than cross-correlation with a photometric sample. Their analysis considered luminosity and color rather than stellar mass and color, and used an $r-i$ rather than $g-i$ color definition, so we cannot make an exact comparison. However, our results are in good agreement in terms of general trends. Guo et al. find that more luminous and redder galaxies generally exhibit stronger clustering, as well as a clear color dependence within the red sequence. At a fixed luminosity, they find that progressively redder galaxies are more strongly clustered on small scales, analogous to our fixed stellar mass result.

\section{Conclusions} \label{sec:conclusions}
We have presented detailed measurements of the small-scale clustering of massive BOSS galaxies as a function of stellar mass and galaxy color, using cross-correlations to a larger set of fainter galaxies. We find that red galaxies at $0.6 < z < 0.65$ have denser environments than their blue counterparts and that this behavior heightens at smaller radii. In the 125--250 kpc annulus, we find the ratio of the clustering amplitudes $w_\text{red}/w_\text{blue}$ to be $1.92 \pm 0.11$. This ratio decreases with increasing radius, with $w_\text{red}/w_\text{blue} = 1.24 \pm 0.05$ in the 2--4 Mpc annulus. That red galaxies have denser environments at $z \sim 0.625$ implies that massive halos (which exhibit denser clustering) have red central galaxies. The opposite appears to be true for blue galaxies: they are predominantly found in lower mass halos, which promote lower clustering density. This has implications for theories of halo formation and evolution and their relationships to galaxy properties.

We were able to obtain very precise measurements of the clustering in each subpopulation. Figure \ref{rw_v_r} clearly shows that the clustering amplitude deviates from a power-law relation for both red and blue galaxies, which is in agreement with results using earlier SDSS data \citep{Zehavi:2004}. The dip around 1 Mpc can be explained by a change in environment from one- to two-halo systems, where the dominant galaxy pairs shift from being those in the same dark matter halo to those in separate halos.  

Isolating stellar mass into smaller bins, we find a clear differentiation between clustering amplitude in red and blue galaxies (especially at smaller scales), showing that stellar mass is not the sole determinant of clustering. This rejects a simple model that halo mass begets stellar mass, independent of color. Holding color fixed, we find that clustering amplitude increases with stellar mass, especially for red galaxies at small radii (where it is more than a factor of 2 effect over 0.75 dex in stellar mass).

For further study, we can divide up the imaging by color so as to study galactic conformity \citep[e.g.,][]{Weinmann:2006, Kauffmann:2013, Hartley:2015, Knobel:2015}. We also look forward to larger data sets with eBOSS \citep{Zhao:2016}, the Subaru Prime Focus Spectrograph \citep[PFS;][]{Takada:2014}, DESI \citep{Levi:2013}, Euclid \citep{Laureijs:2009, Laureijs:2011}, LSST \citep{Tyson:2002}, and WFIRST \citep{Spergel:2013b, Spergel:2013a} that can be matched up to deeper imaging.

\acknowledgements
We are grateful to Aaron Bray for both analysis and editorial help. We thank Ashley Ross for providing the imaging mask, and the SDSS-III BOSS Galaxy Clustering Working Group for providing the catalog of spectroscopic galaxies.

\bibliography{refs}
\bibliographystyle{aasjournal}

\end{document}